\documentclass[conference]{IEEEtran}
\IEEEoverridecommandlockouts

\usepackage{algorithm,algorithmic}
\usepackage{amsmath,amssymb,amsfonts}
\usepackage{algorithmic}
\usepackage{graphicx}
\usepackage{textcomp}
\usepackage{bm}
\usepackage{xcolor}
\usepackage{hyperref}
\usepackage{tikz}
\usetikzlibrary{arrows.meta, positioning, quotes}
\usepackage{booktabs}
\usepackage[skip=2pt]{caption}
\usepackage{subcaption}
\usepackage{nicefrac}
\usepackage{multicol, multirow}
\usepackage{adjustbox}
\usepackage{etoolbox}
\makeatletter
\patchcmd{\hyper@makecurrent}{%
    \ifx\Hy@param\Hy@chapterstring
        \let\Hy@param\Hy@chapapp
    \fi
}{%
    \iftoggle{inappendix}{%true-branch
        \@checkappendixparam{chapter}%
        \@checkappendixparam{section}%
        \@checkappendixparam{subsection}%
        \@checkappendixparam{subsubsection}%
        \@checkappendixparam{paragraph}%
        \@checkappendixparam{subparagraph}%
    }{}%
}{}{\errmessage{failed to patch}}

\newcommand*{\@checkappendixparam}[1]{%
    \def\@checkappendixparamtmp{#1}%
    \ifx\Hy@param\@checkappendixparamtmp
        \let\Hy@param\Hy@appendixstring
    \fi
}
\makeatletter

\newtoggle{inappendix}
\togglefalse{inappendix}

\apptocmd{\appendix}{\toggletrue{inappendix}}{}{\errmessage{failed to patch}}
\def\eqref#1{equation~\ref{#1}}
\def\1{\bm{1}}

\DeclareMathAlphabet{\mathsfit}{\encodingdefault}{\sfdefault}{m}{sl}
\SetMathAlphabet{\mathsfit}{bold}{\encodingdefault}{\sfdefault}{bx}{n}

\DeclareMathOperator{\argmaxinline}{arg\,max}

\newcommand{\matr}[1]{\bm{#1}}
\newcommand{\vect}[1]{\mathbf{#1}}
\newcommand{\vectsym}[1]{\bm{#1}}
\newcommand{\set}[1]{\mathcal{#1}}
\newcommand{\norm}[1]{\left\lVert#1\right\rVert}

\usepackage[natbib=true, sorting=none]{biblatex}
\usepackage[capitalize]{cleveref}

\usepackage{mathtools}
\usepackage{amsthm}
\theoremstyle{plain}
\newtheorem{theorem}{Theorem}[section]
\newtheorem{lemma}{Lemma}[section]
\newtheorem{definition}{Definition}[section]

\def\BibTeX{{\rm B\kern-.05em{\sc i\kern-.025em b}\kern-.08em
    T\kern-.1667em\lower.7ex\hbox{E}\kern-.125emX}}

\title{Efficient MILP Decomposition in Quantum Computing for ReLU Network Robustness
\thanks{The project/research is supported by the Bavarian Ministry of Economic Affairs, Regional Development and Energy with funds from the Hightech Agenda Bayern.}
}

\usepackage{fancyhdr}

\fancypagestyle{specialfooter}{%
  \fancyhf{}
  
  \fancyfoot[R]{ \noindent\fbox{%
    \parbox{\textwidth}{%
        {\footnotesize \copyright 2023 IEEE. Personal use of this material is permitted. Permission from IEEE must be obtained for all other uses, in any current or future media, including reprinting/republishing this material for advertising or promotional purposes, creating new collective works, for resale or redistribution to servers or lists, or reuse of any copyrighted component of this work in other works.}
        }
    }}
}

\author{
    \IEEEauthorblockN{Nicola Franco\IEEEauthorrefmark{2}, Tom Wollschl\"ager\IEEEauthorrefmark{3}, Benedikt Poggel\IEEEauthorrefmark{2}, Stephan G\"unnemann\IEEEauthorrefmark{3}, Jeanette Miriam Lorenz\IEEEauthorrefmark{2}}
    \IEEEauthorblockA{\IEEEauthorrefmark{2}Fraunhofer Institute for Cognitive Systems IKS, Munich, Germany
    \\\{nicola.franco, benedikt.poggel, jeanette.miriam.lorenz\}@iks.fraunhofer.de}
    \IEEEauthorblockA{\IEEEauthorrefmark{3}Dept. of Computer Science \& Munich Data Science Institute, Technical Univ. of Munich, Germany
    \\\{tom.wollschlaeger, s.guennemann\}@tum.de}
}

\begin{document}

\maketitle
\thispagestyle{plain}
\pagestyle{plain}
\thispagestyle{specialfooter} % Only for the pre-print

\begin{abstract}
    Emerging quantum computing technologies, such as Noisy Intermediate-Scale Quantum (NISQ) devices, offer potential advancements in solving mathematical optimization problems. 
    However, limitations in qubit availability, noise, and errors pose challenges for practical implementation. 
    In this study, we examine two decomposition methods for Mixed-Integer Linear Programming (MILP) designed to reduce the original problem size and utilize available NISQ devices more efficiently.
    We concentrate on breaking down the original problem into smaller subproblems, which are then solved iteratively using a combined quantum-classical hardware approach. 
    We conduct a detailed analysis for the decomposition of MILP with Benders and Dantzig-Wolfe methods.
    In our analysis, we show that the number of qubits required to solve Benders is exponentially large in the worst-case, while remains constant for Dantzig-Wolfe.
    Additionally, we leverage Dantzig-Wolfe decomposition on the use-case of certifying the robustness of ReLU networks. 
    Our experimental results demonstrate that this approach can save up to 90\% of qubits compared to existing methods on quantum annealing and gate-based quantum computers.
\end{abstract}

\begin{IEEEkeywords}
Quantum Computing, Mixed-Integer Linear Programming, Hybrid Algorithm
\end{IEEEkeywords}

\section{Introduction}\label{sec:intro}

In recent years, remarkable progress has been made in the field of Quantum Computing (QC) in terms of both hardware and software development. This includes the experimental demonstration of quantum error correction, which starts to enhance performance as qubit count increases~\cite{google2023suppressing}. These advancements have broadened the practical capabilities of Noisy Intermediate-Scale Quantum (NISQ) devices, allowing them to tackle more complex challenges.

Operations research, with its wide-ranging real-world applications across finance, logistics, manufacturing, and automotive industries, has emerged as a particularly promising area for NISQ devices. \textit{Mixed-integer linear programming} (MILP), a common problem formulation in operations research, involves a combination of integer and continuous variables constrained by linear equations.
MILPs frequently involve complex combinatorial optimization problems that pose difficulties for classical solvers, particularly when dealing with large-scale instances, as they are NP-hard~\cite{papadimitriou1982complexity}. 
As such, QC holds the potential to significantly accelerate the solving process and enhance overall efficiency in addressing these problems~\cite{grover1997quantum}.
Recent works shows that approximation~\cite{montanez2022unbalanced}, reduction~\cite{ranvcic2023noisy} and decomposition~\cite{chang2020quantum, gambella2020multiblock, zhao2022hybrid} approaches are needed towards the possibility of gaining potential advantages with QC.
In the context of MILP, two decomposition methods have shown some potential: Benders~\cite{benders1962partitioning} and Dantzig-Wolfe~\cite{dantzig1960decomposition}.
The objective of both approaches is to break down the original problem into smaller instances to enable more efficient use of quantum computing for large-scale optimization problems.
\\
\newline
QC and MILP are not only revolutionizing operations research but also presenting promising opportunities for formal verification of neural networks. 
As neural networks become increasingly prevalent, ensuring their reliability, robustness, and security through formal verification is crucial, especially in safety-critical applications.
In this context, formal verification of neural networks aims to provide mathematical guarantees of their expected behavior under predefined conditions.
This often involves proving properties such as robustness against adversarial attacks, generalization, and compliance with safety constraints. 
One prevalent technique transforms the verification problem into a MILP problem, solvable using existing solvers~\cite{katz2017reluplex, tjeng2018evaluating}. 
However, the exact MILP solution is computationally challenging due to its exponential complexity, specifically for large networks.
As a result, researchers are motivated to investigate quantum optimization algorithms as an alternative solution approach~\cite{franco2022quantum}.
\\
\newline
In this work, we compare Benders and Dantzig-Wolfe decompositions for MILP in terms of complexity and qubits requirements.
Particularly, one application we explore involves using QC to verify the robustness of Rectified Linear Unit (ReLU) networks.
Since ReLU non-linearity can be reformulated as a binary variable, the verification problem can be represented as a MILP~\cite{katz2017reluplex, ehlers2017formal, tjeng2018evaluating}.
Expanding on our previous approach \cite{franco2022quantum}, we propose a Hybrid Quantum-Classical Robustness Analyzer for Neural Networks with Dantzig-Wolfe decomposition (HQ-CRAN-DW). 
This method iteratively addresses the MILP formulation using a combination of classical and quantum hardware. 
In contrast to previous studies, our approach adopts the Dantzig-Wolfe reformulation of the initial problem, providing a close representation in terms of the dual. 
It is essential to recognize that this method relies on a linear programming relaxation, which introduces a limitation on the tightness of the original problem. 
Despite this, the key advantage of HQ-CRAN-DW is evident in the reduced number of qubits required when transitioning from a constrained to an unconstrained problem.

The contributions of our work are: 
\begin{itemize}
    \item Analyzing the decomposition of MILPs with Benders~\cite{benders1962partitioning} and Dantzig-Wolfe~\cite{dantzig1960decomposition} for QC in terms of complexity and qubits requirements.
    \item Demonstrating that the number of qubits required to solve Benders is exponentially large in the worst-case, while it remains constant for Dantzig-Wolfe.
    \item Leveraging Dantzig-Wolfe decomposition on the use-case of certifying the robustness of ReLU networks.
    \item Achieving up to a 90\% reduction in qubit usage compared to existing methods on quantum annealing and gate-based quantum computers.
\end{itemize}
Code is available at \href{https://github.com/FraunhoferIKS/hqcran}{https://github.com/FraunhoferIKS/hqcran}
\section{Related Works}\label{sec:related}

A variety of quantum optimization algorithms have recently been suggested for use with NISQ devices, aiming to address large-scale combinatorial optimization problems that are typically difficult for classical solvers to handle.
Many optimization problems can be mapped into Quadratic Unconstrained Binary Optimization (QUBO) form, making it a versatile and convenient framework for leveraging the potential power of quantum computing.
QUBO problems can then be optimized directly using widely-used variational quantum algorithms, such as the Variational Quantum Eigensolver (VQE)~\cite{peruzzo2014variational} or the Quantum Approximate Optimization Algorithm (QAOA)~\cite{farhi2014quantum}.
Generally, in order to transform a MILP problem into a QUBO, real variables must be approximated as binary variables.

Although it is possible to apply \citet{grover1997quantum} search to optimization problems directly on real variables~\cite{protopopescu2002solving}, implementing it on actual hardware is not practical for large-scale problems on NISQ devices due to the high gate complexity involved.
Instead, potential approaches for solving large-scale problems on NISQ devices include using slack-based formulations and treating the slacks as extra continuous parameters for quantum QUBO solvers~\cite{braine2021quantum}. 
A second approach is to utilize a fixed-point approximation with binary variables, as described by~\citet{vyskovcil2019embedding}. 
A third approach is to substitute the slack variables with a fixed choice of hyper-parameters to the first and second-order Taylor expansion of the constraints, as proposed by~\citet{montanez2022unbalanced}. 
These approaches offer practical solutions for implementing optimization problems on NISQ devices.

In the context of mixed-integer problems, \citet{gambella2020multiblock} introduced a decomposition technique based on the alternating direction method of multipliers (ADMM), which heuristically solve mixed-binary optimization problems.
Newer methods focus on utilizing decomposition strategies such as Benders~\cite{chang2020quantum, zhao2022hybrid, franco2022quantum} or Dantzig-Wolfe~\cite{ossorio2022optimization, coelho2023quantum}.
However, there is no clear comparison of which method is best suited for QC-based MILP optimization.
To address this gap, this study provides an overview of the qubit requirements and complexity of these two decomposition methods, enabling a comparison of their suitability for QC-based MILP optimization.

\section{Preliminaries}\label{sec:preliminaries}

MILPs represent a class of problems with continuous and integer variables where the objective function and the constraints are linear.
A MILP in its \textit{canonical form} is expressed through:

\begin{subequations}\label{eq:canonical_form}
    \begin{align}
        \min_{\vect{x}, \vect{y}} \quad &\vect{c}^\intercal \vect{x} + \vect{d}^\intercal \vect{y}, \label{eq:canonical_form_objective} \\
        \text{s.t.}\quad &\matr{A}\vect{x} + \vect{B}\vect{y} \geq \vect{b}, \label{eq:canonical_form_complicating} \\
        &\vect{x} \in \set{X},\; \vect{y} \in \set{Y} \label{eq:canonical_form_easy},
    \end{align}
\end{subequations}
where $\vect{c}\in\mathbb{Q}^{n_x}$, $\vect{d}\in\mathbb{Q}^{n_y}$, $\vect{b}\in\mathbb{Q}^{m}$ are vectors and $\matr{A}\in\mathbb{Q}^{m\times n_x}$, $\matr{B}\in\mathbb{Q}^{m\times n_y}$ are matrices.
Additionally, we denote (\ref{eq:canonical_form_complicating}) as \textit{complicating} constraints, while we define (\ref{eq:canonical_form_easy}) as the set of \textit{easy} constraints, where $\set{X} \subseteq \mathbb{R}^{n_x}$  and $\set{Y} \subseteq \mathbb{Z}^{n_y}$ are polyhedra for real and integer variables, respectively\footnote{e.g. $\set{X} = \{\vect{x} \in \mathbb{R}^{n_x}\;\colon \matr{C}\vect{x} \geq \vect{d}\}$ and $\set{Y} = \{\vect{y} \in \mathbb{Z}^{n_y} \;\colon \matr{E}\vect{y} \geq \vect{g} \}$.}.

As of now, MILPs cannot be directly solved using Variational Quantum Algorithms such as VQE or QAOA, as these algorithms are tailored to optimize QUBO formulations.
Therefore, decomposition methods like Benders~\cite{benders1962partitioning} and Dantzig-Wolfe~\cite{dantzig1960decomposition} are essential for breaking down MILPs into smaller, more manageable subproblems, which can then be transformed into QUBO representations compatible with quantum optimization algorithms. 

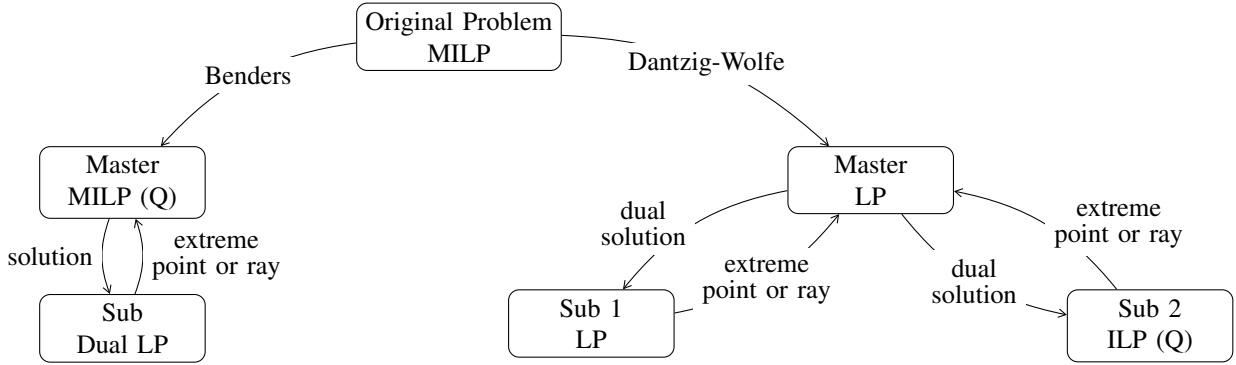
\begin{figure*}[ht]
    \vspace{-1em}
    \centering
    \begin{tikzpicture}[
        node distance=2cm,
        box/.style = {draw, rounded corners, 
                     minimum width=22mm, minimum height=5mm, align=center},
        > = {Straight Barb[angle=60:2pt 3]},
        bend angle = 20,
        ]
        
        \node (n1)  [box] {Original Problem \\ MILP};
        \node (n2)  [box, below left=1cm and 2cm of n1] {Master\\ MILP (Q)};
        \node (n3)  [box, below=1cm of n2]    {Sub\\ Dual LP};
        \node (n4)  [box, below right=1cm and 3cm of n1]    {Master\\ LP};
        \node (n5)  [box, below left=1cm and 1.5cm of n4]    {Sub 1\\ LP};
        \node (n6)  [box, below right=1cm and 1.5cm of n4]    {Sub 2\\ ILP (Q)}; 
    
        \draw[->]   (n1) edge[bend right] node[midway, color=black, fill=white]{Benders}          (n2);
        \draw[->]   (n1) edge[bend left] node[midway, color=black, fill=white]{Dantzig-Wolfe}     (n4);
        \draw[->]   (n2) edge[bend right] node[left, color=black, fill=white]{solution}           (n3);
        \draw[->]   (n3) edge[bend right] node[right, color=black, fill=white]{\shortstack{extreme\\point or ray}}     (n2);
        \draw[->]   (n4) edge[bend right] node[left, color=black, fill=white]{\shortstack{dual\\ solution}}    (n5);
        \draw[->]   (n5) edge[bend right] node[midway, color=black, fill=white]{\shortstack{extreme\\point or ray}}    (n4);
        \draw[->]   (n6) edge[bend right] node[right, color=black, fill=white]{\shortstack{extreme\\point or ray}}    (n4);
        \draw[->]   (n4) edge[bend right] node[midway, color=black, fill=white]{\shortstack{dual\\ solution}}    (n6);
    
    \end{tikzpicture}
    \caption{This diagram offers an high-level overview of Benders and Dantzig-Wolfe decomposition for a MILP.
    We identify with (Q) the problems that are optimized with quantum hardware.}
    \label{fig:diagram}
    \vspace{-1em}
\end{figure*}

In \autoref{fig:diagram}, we offer a high-level summary of both techniques. 
Linear programming (LP) refers to problem instances with continuous variables and linear constraints, whereas integer linear programming (ILP) pertains to problems with integer variables and linear constraints. 
Additionally, we label problems transformed into QUBO format and solved using QC with (Q).

\subsection{Benders Decomposition}\label{sec:benders}

Benders decomposition~\cite{benders1962partitioning} splits the original problem into two subsets of variables: A mixed-integer set and a linear set (real). 
The mixed-integer set is the \textit{master} problem and is solved using the initial set of variables, while the second set of variables is determined in a \textit{subproblem}, given a master solution. 
If the subproblem finds the fixed master decision to be infeasible, valid inequalities known as Benders cuts, are generated and incorporated into the master problem, which is then solved again until no more cuts can be produced.

Let us rewrite \cref{eq:canonical_form} as $\textstyle\min_{\vect{y} \in \set{Y}} \vect{d}^\intercal \vect{y} + q(\vect{y})$ where: 
\begin{equation}\label{eq:reformulated}
    q(\vect{y}) = \min_{\vect{x}\in\set{X}} \left\{\vect{c}^{\intercal}\vect{x}
   \;\colon \matr{A}\vect{x} \geq \vect{b} - \matr{B}\vect{y}\right\}.
\end{equation}
Here we view the vector of binary variables $\vect y$ as given. 
Hence, we decoupled $\vect y$ from the rest of the program resulting in a LP. 
Thus, we can consider the dual formulation of $q(\vect{y})$ as:
\begin{equation}\label{eq:dual}
    \max_{\vectsym{\alpha}\in \mathbb{R}^{m_b}_+} \; \left\{\vectsym{\alpha}\left( \vect{b} - \vect{B} \vect{y} \right) \;\colon \vectsym{\alpha} \matr{A} = \vect{c}^\intercal \right\},
\end{equation}
where w.l.o.g. and for ease of exposition we consider $\vectsym{\alpha}\in \mathbb{R}^{m_b}_+$ as row vector and $\set{X} = \mathbb{R}^{n_x}$.
Since $\vect{y}$ is constant within the optimization of \cref{eq:reformulated}, the optimization program is a LP and we thus have strong duality\footnote{i.e. the optimal objective value of the primal equals the optimal value of the dual.}. 
%The optimal objective value of \cref{eq:dual} is infinity if \autoref{eq:reformulated} is infeasible. If \autoref{eq:reformulated} is feasible it has to have the same finite value. Thus, it should be either finite or positive infinity. %\tom{do we need to cite chang here again?}
%Therefore, for an optimal value of the dual formulation of \autoref{eq:dual}, $\vect{g}^\intercal - \vectsym{\alpha}\matr{A} - \vectsym{\beta}^\intercal \matr{C}$ has to be zero. 
%because otherwise, taking the infimum over all $\vect{z}$ will result in negative infinity. 
%Hence, we know the optimal feasible objective will only be the result of the remaining terms that are not interacting with $\vect{z}$ as all of them need to cancel out. 
% So we can reformulate the dual as\footnote{See \citet{chang2020quantum} for more details.}:
Following Benders decomposition~\cite{benders1962partitioning, geoffrion1972generalized}, we can formulate the objective of \cref{eq:dual} as a linear combination of extreme rays and points of the feasible region. We denote as $\Lambda_r$ and $\Lambda_p$ the set of extreme rays and extreme points of the set $\left\{ \vectsym{\alpha} \in \mathbb{R}_{+}^{m_b} \;\colon  \vectsym{\alpha} \matr{A} = \vect{c}^\intercal \right\}$.

Hence, we can describe $q(\vect{y})$ in \cref{eq:canonical_form} in terms of extreme points and rays as an \textit{exponential} set of cuts to obtain the following representation:
\begin{subequations} \label{eq:master_problem}
\begin{alignat}{2} 
    &\min_{\vect{y},\, \eta} \, \vect{d}^\intercal \vect{y} + \eta, \\
    \text{s.t.} \
    &\vectsym{\alpha}^{(k)}\left( \vect{b} - \matr{B} \vect{y} \right) \leq \eta, \; \forall \vectsym{\alpha}^{(k)} \in \Lambda_p, \label{eq:master_extreme_point}\\
    &\vectsym{\alpha}^{(k)}\left( \vect{b} - \matr{B} \vect{y} \right) \leq 0, \; \forall \vectsym{\alpha}^{(k)} \in \Lambda_r, \label{eq:master_extreme_ray}
\end{alignat}
\end{subequations}
where $\eta\in\mathbb{R}$ is a scalar. 
This problem is known as \textit{master} problem on Benders decomposition. 

As noted, the difficulty of solving \cref{eq:master_problem} is the exponential size of the sets $\Lambda_p, \Lambda_r$.
% \cite{chang2020quantum}. 
Thus, we can gradually extend the sets $\Lambda^\prime_p \subseteq \Lambda_p, \Lambda^\prime_r \subseteq \Lambda_r$ by constraints of the \textit{subproblem} defined as:
\begin{equation} \label{eq:sub_problem}
        \max_{\vectsym{\alpha} \leq \bar{\vectsym{\alpha}}}
        \left\{\vectsym{\alpha}\left( \vect{b} - \matr{B} \vect{y} \right)\;\colon
        \vectsym{\alpha} \matr{A} = \vect{c}^\intercal \right\}.
\end{equation}
The subproblem is similar to \cref{eq:dual} except that $\vectsym{\alpha}$ is bounded.
This is an iterative procedure known as \textit{delayed constraint generation}, where at every step $t$ there exists three possible outcomes: (i) if the solution of \cref{eq:sub_problem} is unbounded ($\exists \vectsym{\alpha}_i \in  \vectsym{\alpha}^{(t)}\; \colon \vectsym{\alpha}_i = \bar{\vectsym{\alpha}}_i$), then we obtain an extreme ray $\Lambda_r^\prime \leftarrow \vectsym{\alpha}^{(t)}$; (ii) if the solution of \cref{eq:sub_problem} is lower than the solution of \cref{eq:master_problem}, then we acquire an extreme point $\Lambda_p^\prime \leftarrow \vectsym{\alpha}^{(t)}$; (iii) if the solution of \cref{eq:sub_problem} is equal to the solution of \cref{eq:master_problem}, the algorithm terminates. 
In practice, we define a threshold quantity that allows us to stop the algorithm when the two solutions are sufficiently close to each other.

\subsection{Dantzig-Wolfe Decomposition}\label{sec:dantzig-wolfe}

Dantzig–Wolfe decomposition is an algorithm for solving linear programming problems with special structure\footnote{a block-angular or block-diagonal arrangement in the constraint matrix.}. 
This decomposition relies on a delayed column generation for improving the tractability of large-scale linear programs. 
For MILP problems solved via the Dantzig-Wolfe, at each step, most columns (variables) are not in the \textit{basis}.
In this context, a \textit{basis} refers to a collection of linearly independent columns from the constraint matrix, which form the current active solution set.
In such a scheme, a master problem containing at least the currently active columns (the basis) uses a subproblem or subproblems to generate columns for entry into the basis such that their inclusion improves the objective function.
The traditional decomposition method relies on Minkowski and Weyl's theorem~\cite{schrijver1998theory}, which serves as its foundation, and for MILP employs a \textit{convexification} process.
Let $\set{U}$ be the feasible region of \cref{eq:canonical_form}.
\begin{definition}[feasible region]\label{def:feasible_region}
     A feasible region $\set{U}$ is the set of all possible points of \cref{eq:canonical_form} that satisfy the problem's constraints:
     \begin{equation}
        \set{U} = \left\{\vect{x}\in\set{X}, \vect{y}\in\set{Y}\;\colon \matr{A}\vect{x} + \matr{B}\vect{y} \geq \vect{b} \right\}.
    \end{equation}
\end{definition}

Minkowski and Weyl's theorem~\cite{schrijver1998theory} states that every polyhedron $\set{U}$ can be written as 
sum of finitely many extreme points and extreme rays.
Thus, we denote its sets of extreme points with $\set{P}_\set{X}=\left\{ \vect{x}^{(i)},\; \forall i \in \set{I} \right\}$ and $\set{P}_\set{Y}=\left\{ \vect{y}^{(j)},\; \forall j \in \set{J} \right\}$\footnote{For the sake of readability we only consider extreme points and not extreme rays.}.
This allow us to express \cref{eq:canonical_form} as linear combination of its extreme points:
\begin{subequations}\label{eq:dantzig-wolfe-plain}
    \begin{align}
    \min_{\scriptsize \begin{array}{c} \lambda_i, \forall i \in I\\ \mu_j \forall j \in J \end{array}}\; &\sum_{i \in \set{I}}(\vect{c}^\intercal \vect{x}^{(i)})\lambda_i + \sum_{j \in \set{J}} (\vect{d}^\intercal \vect{y}^{(j)}) \mu_j, & \\
    \text{s.t.}\quad &\sum_{i \in \set{I}} (\matr{A} \vect{x}^{(i)})\lambda_i 
        + \sum_{j \in \set{J}} (\matr{B} \vect{y}^{(j)}) \mu_j \geq \vect{b}, \label{eq:dw-coupling} \\
    &\sum_{i \in \set{I}} \lambda_i = 1,\quad \lambda_i \geq 0,\; \forall i \in \set{I}, \label{eq:dw-convexity_real}\\
    &\sum_{j \in \set{J}} \mu_j = 1,\quad \mu_j \geq 0,\; \forall j \in \set{J}, \label{eq:dw-convexity_binary}\\
    &\vect{y} = \sum_{j\in\mathcal{J}}\vect{y}^{(j)}\mu_j, \quad \vect{y} \in \mathbb{Z}^{n_y}, \label{eq:dw-integrality}
    \end{align}
\end{subequations}
where the variables $\lambda_i \in \mathbb{R}$ and $\mu_j \in \mathbb{R}$ represent the weights of each extreme point for real and integer variables, respectively.
In this context, \cref{eq:dantzig-wolfe-plain} is typically called the \textit{master} problem. 
In addition, constraint (\ref{eq:dw-coupling}) is denoted as the \textit{coupling} constraint and constraints (\ref{eq:dw-convexity_real}-\ref{eq:dw-convexity_binary}) are called \textit{convexity} constraints.
It is important to note that integrality is still imposed on the original $\vect{y}$ variable through \cref{eq:dw-integrality}.

This representation of Dantzig-Wolfe decomposition is known as \textit{convexification} approach and may not be straightforward in general~\cite{desrosiers2010branch}. 
However, in the significant special case of combinatorial optimization with QC where $\mathcal{Y}$ is a subset of $\{0, 1\}^{n_y}$, \textit{convexification} and \textit{discretization} coincide~\cite{vanderbeck2010reformulation}\footnote{In this work, we only consider the convexification approach for the sake of conciseness.}. 
Additionally, both techniques produce the same dual bound, which is equal to that of Lagrangean relaxation~\cite{geoffrion1974lagrangian}.

Since $\set{I}$ and $\set{J}$ contain an exponential number of extreme points,  \cref{eq:dantzig-wolfe-plain} will have an exponential number of variables compared to \cref{eq:canonical_form}.
Thus, we consider a restricted version of \cref{eq:dantzig-wolfe-plain} by progressively adding each new extreme point to the subsets $\set{I}^\prime \subseteq \set{I}$ and $\set{J}^\prime \subseteq \set{J}$.
To determine which extreme point to include, we define two subproblems, referred to as \textit{pricing} problems, by considering the dual of \cref{eq:dantzig-wolfe-plain}.

\subsubsection{Dual formulation of \cref{eq:dantzig-wolfe-plain}}\label{sec:dual}

In accordance with the work of \cite{geoffrion1974lagrangian, vanderbeck2010reformulation}, we transition to the dual representation of \cref{eq:dantzig-wolfe-plain}.
Thus, let us introduce the so called \textit{Lagrangean} subproblem $\max_{\vectsym{\alpha}, \xi, \eta} \mathcal{L}(\vectsym{\alpha}, \xi, \eta)$, with $\mathcal{L}$ defined as:
\begin{equation}
\begin{aligned}
    \min_{\scriptsize \begin{array}{c} \lambda_i, \forall i \in \set{I^\prime}\\ \mu_j \forall j \in \set{J^\prime} \end{array}} &\sum_{i \in \set{I}}  \vectsym{\alpha}\vect{b} + \left( (\vect{c}^\intercal - \vectsym{\alpha} \matr{A})\vect{x}^{(i)} + \xi \right) \lambda_i - \xi \\
    &+ \sum_{j \in \set{J}} \left( (\vect{d}^\intercal - \vectsym{\alpha} \matr{B})\vect{y}^{(j)} + \eta \right)\mu_j -\eta,
\end{aligned}
\end{equation}
where $\vectsym{\alpha}\in\mathbb{R}^m_+$ is a row vector and $\xi, \eta \in\mathbb{R}$ are scalars (also known as Lagrangian multipliers).
It is worth noting that we omit the integrality constraint from \cref{eq:dw-integrality} in the reformulation, as it can be violated at a price of $\vectsym{\alpha}$~\cite{vanderbeck2010reformulation}.
The process of raising an integer (or mixed-integer) problem to a higher-dimensional space, deriving an enhanced formulation in that context, and subsequently returning it to the initial variable space is a familiar strategy in integer programming~\cite{desrosiers2010branch, vanderbeck2010reformulation}.

The solution of the function $\mathcal{L}$ establishes a dual (lower) bound on the optimal value of \cref{eq:dantzig-wolfe-plain}. 
The task of maximizing this bound across the set of acceptable penalty vectors is referred to as the Lagrangean dual:
\begin{equation}\label{eq:dual-master-dantzig-wolfe}
\begin{aligned}
    \max_{\vectsym{\alpha}, \xi, \eta} \quad &\vectsym{\alpha} \vect{b} -\xi - \eta, \\
    \text{s.t.} \quad 
    &(\vect{c}^\intercal - \vectsym{\alpha} \matr{A})\vect{x}^{(i)} + \xi \leq 0, \quad \forall i \in \set{I}, \\ 
    & (\vect{d}^\intercal - \vectsym{\alpha} \matr{B})\vect{y}^{(j)} + \eta  \leq 0, \quad \forall j \in \mathcal{J}, \\ 
\end{aligned}
\end{equation}
which is known as \textit{dual master} problem on Dantzig-Wolfe decomposition.
\begin{lemma}[Lagrangian bound~\cite{vanderbeck2006generic}]\label{def:lagrangian_dual}
    The solution of \cref{eq:dantzig-wolfe-plain} offers a dual bound that is equal to the solution of \cref{eq:dual-master-dantzig-wolfe}.
\end{lemma}

\begin{table*}[ht]
    \vspace{-1em}
    \centering
    \caption{A comparative summary of MILP decomposition methods for quantum computing, detailing the complexity of the master and subproblems (P: polynomial-time solvable, NP-hard: non-deterministic polynomial-time hard), along with the number of qubits required at the first and $2^{n_y}$-th iterations. The terms $n_s$ and $m_y$ represent the number of slack variables and the number of constraints involving integer variables, respectively.}
    \label{tab:decompostion_methods}
    % \begin{adjustbox}{width=0.7\textwidth}
    \begin{tabular}{lllll}
    \toprule
    \multirow{2}{*}{\textbf{Method}} &\multicolumn{2}{c}{\textbf{Complexity}} &\multicolumn{2}{c}{\textbf{\# of qubits at iteration}}\\
    &\multicolumn{1}{c}{\textbf{Master}} &\multicolumn{1}{c}{\textbf{Sub}} &\multicolumn{1}{c}{\textbf{1st}} &\multicolumn{1}{c}{\textbf{2\textsuperscript{$n_y$}-th}} \\
    \midrule
    Benders         &NP-hard (QUBO) &P (Dual LP)  &$\mathcal{O}(n_y + 2\cdot n_s)$     &$\mathcal{O}(n_y + 2^{n_y}\cdot n_s)$ \\
    Dantzig-Wolfe   &P (LP)    &NP-hard (QUBO)    &$\mathcal{O}(n_y + m_y \cdot n_s)$  &$\mathcal{O}(n_y + m_y \cdot n_s)$ \\
    \bottomrule
    \end{tabular}
    % \end{adjustbox}
    \vspace{-1em}
\end{table*}

\subsubsection{Column generation}

Dantzig-Wolfe decomposition involves iterating between the master and subproblems, which are also called \textit{pricing problems}.
This method is commonly referred to as the \textit{column generation process}~\cite{dantzig1960decomposition}.
To initiate the process, a preliminary restricted master problem is required.
Having a feasible linear programming relaxation for this initial restricted master problem is essential, as it ensures the proper exchange of dual information with the pricing problems.
At every step $t$, we generate an extreme point $\vect{x}^{(t)}$, and an extreme point $\vect{y}^{(t)}$.
These extreme points are incorporated into the master, necessitating the addition of new $\lambda_i$ and $\mu_j$ columns.
The \textit{real pricing} problem is given as: 
\begin{equation}\label{eq:sub-1-dantzig-wolfe}
    \min_{\vect{x}\in\set{X}}\; (\vect{c}^\intercal - \vectsym{\alpha}^{{(t)}} \matr{A})\vect{x},
\end{equation}
where $\vectsym{\alpha}^{(t)}$ is the dual solution of \cref{eq:dantzig-wolfe-plain} associated with the constraint (\ref{eq:dw-coupling}).
Similarly, the \textit{integer pricing} problem is given as:
\begin{equation}\label{eq:sub-2-dantzig-wolfe}
    \min_{\vect{y}\in\set{Y}}\; (\vect{d}^\intercal -\vectsym{\alpha}^{{(t)}} \matr{B})\vect{y},
\end{equation}
%As one can see, the solution of \cref{eq:dual-master-dantzig-wolfe} directly affects the objective of both sub problems.
which deals with integer variables and therefore has a stronger complexity.
If the solution of \cref{eq:sub-1-dantzig-wolfe} is lower then $\xi$, then we set $\set{I}^\prime \leftarrow \vect{x}^{(t)}$. 
Similarly, if the solution of \cref{eq:sub-2-dantzig-wolfe} is lower than $\eta$, than we add $\set{J}^\prime \leftarrow \vect{y}^{(t)}$.
Analogously to Benders, we define a threshold quantity $\theta$ that allows us to stop the algorithm when the difference between the solution of \cref{eq:dantzig-wolfe-plain} and its dual is lower than $\theta$.

\subsection{Quadratic Unconstrained Binary Formulation}\label{sec:qubo}

By transforming Benders master problem or Dantzig-Wolfe integer pricing problem into a QUBO problem, the power of quantum optimization algorithms, such as VQE or QAOA, can be harnessed to find more efficient solutions. 
The transformation involves rewriting the objective function and constraints of each subproblem using binary variables, and then converting them into a quadratic cost function in line with the QUBO formulation:
\begin{equation}
\min_{\vect{q} \in \{0, 1\}^{n_q}} \vect{q}^\intercal \matr{Q} \vect{q},    
\end{equation}
where $\matr{Q}\in\mathbb{R}^{n_q \times n_q}$. QUBO problems can be directly converted to an Ising model and vice versa~\cite{barahona1989experiments}, which is the reason for its use in QC.
\section{Benders vs. Dantzig-Wolfe in QC-based MILP Solving}\label{sec:comparison}

In this section, we compare the two previously presented decomposition methods in terms of qubits requirements and complexity. 
To recap, while both Benders and Dantzig-Wolfe techniques aim to break down MILP problems into smaller components to solve them more efficiently, they apply to different types of problem structures and utilize different strategies. 
Benders decomposition is more suitable for problems with a clear separation of integer and continuous variables, while Dantzig-Wolfe decomposition is best for problems with a block structure in the constraints.
In the context of linear programming, it is important to remind that Dantzig-Wolfe decomposition in the primal problem is equivalent to Benders decomposition in the dual problem, with both approaches sharing identical sub-problems~\cite{dantzig2003linear}.
While the two approaches are equivalent, certain stabilization techniques can be more easily formulated in the dual problem compared to the primal problem\footnote{In this context, a primal problem refers to the original optimization problem.}.

In \autoref{tab:decompostion_methods}, we present a comparison between Benders and Dantzig-Wolfe decomposition concerning complexity and the number of qubits needed for solving the QUBO problem with QC. 
A notable advantage of Dantzig-Wolfe over Benders decomposition lies in the fewer qubits required to transform the problem from constrained to unconstrained, which remains constant at each step.

In our comparison, we consider a fixed number of qubits $n_y$ to represent the vector of variables $\vect{y}$. 
Additionally, we consider a fixed number of qubits $n_s$ to convert real variables to binary. 
This consideration is independent of the approximation method used, such as fixed point or floating representation~\cite{vyskovcil2019embedding}.
For example, a fixed-point approximation of a positive real variable is given by $\tilde{\eta} = w\cdot \sum_{i = 0}^{n_s - 1} 2^{i}\cdot y_i$, where $w$ is typically chosen as $10^{-1}$ or $10^{-2}$.
Finally $m_y$ denotes the number of constraints of the set $\set{Y}$.

The difference between the two methods lies in the way on how the problem affected by the QUBO transformation is formulated.
In the context of Benders, the master starts with one cut in the constraints set and a real objective $\eta$.
To convert the problem from constrained to unconstrained an additional slack variable is needed.
Therefore, if we assume the same approximation factor $n_s$ for the slack variable and the real objective $\eta$, at least $2\cdot n_s$ qubits are needed~\cite{chang2020quantum, zhao2022hybrid, franco2022quantum}.
Since, at every step, a new cut is added to the master problem a new slack variable is required.
In the end, if all cuts from the extreme point set are added to the master problem, the number of qubits required by Benders becomes exponential.
In contrast to Dantzig-Wolfe, where the integer pricing problem is fixed in the number of constraints and consequently in the number of binary variables required to approximate the constraints.

It is crucial to highlight that the solution of the master problem in Benders decomposition directly influences the feasibility of the entire problem. 
The heuristic nature of quantum optimization algorithms, such as VQE or QAOA, affects the quality of the solution, which can lead to incorrect cut generation in the corresponding subproblem and ultimately result in infeasible solutions.
On the other hand, a key advantage of the Dantzig-Wolfe decomposition is inherently linked to how the master problem is solved. 
Since the master problem is addressed using classical methods, the coupling constraints condition is consistently satisfied, resulting in more feasible solutions.
Nonetheless, we cannot make the same claim for the individual constraints of the integer pricing problem, as the use of quantum optimization algorithms could still result in unsatisfied constraints in some cases, leading to the generation of false extreme points.
Furthermore, the application of effective heuristics has always been encouraged in the context of Dantzig-Wolfe decomposition to accelerate the overall search process~\cite{desrosiers2010branch}.

In conclusion, in the context of addressing MILP problems with QC, Dantzig-Wolfe decomposition is considered a more favorable choice compared to Benders decomposition.
This preference can be observed in both the constant number of qubits required and the improved feasibility of the resulting solution.
\section{Dantzig-Wolfe for Formal Verification of Neural Network}

In this section, we introduce the task of assessing neural network robustness through the application of Dantzig-Wolfe decomposition. 
We propose a hybrid decomposition method that iteratively solves the MILP formulation by employing both classical and quantum hardware. 
In constrast to \cite{franco2022quantum}, we consider the Dantzig-Wolfe formulation of the original problem, which provides a close representation in terms of the dual.
The main advantages of our approach are demonstrated by the reduced number of qubits needed and the increased number of feasible solutions when utilizing quantum hardware.
We begin by providing a brief overview of the robustness verification problem. 

\subsection{Robustness Certification of Neural Networks}\label{sec:robustness_certificates}

We represent a neural network as a function $\vect{f}(\vect{z}) \colon \set{Z} \to \mathbb{R}^{\lvert \mathcal{K} \rvert}$, which maps input samples $\vect{z} \in \set{Z}$ to output $\vect{k} \in \mathbb{R}^{\lvert \set{K} \rvert}$. 
Here, $\set{K}$ denotes the set of classes. 
We assume a feedforward architecture that consists of affine transformations followed by ReLU activations given as:
\begin{equation}
\begin{aligned}
    \hat{\vect{x}}^{[i]} &= \vect{W}^{[i]}\vect{x}^{[i-1]} + \vect{v}^{[i]},\\
    \vect{x}^{[i]} &= \max{\{0, \hat{\vect{x}}^{[i]}\}},\quad \forall i \in \left\{1,\dots, L\right\},    
\end{aligned}
\end{equation}
where $L$ represents the number of layers, $\vect{x}^{[0]} \equiv \vect{z}$ and $\vect{f}(\vect{z}) \equiv \vect{x}^{[L]}$. 
In case of classification, the network outputs a vector in $\mathbb{R}^{\lvert{\set K}\rvert}$. 
The predicted class is then given by the index of the largest value of that vector, i.e. $c = \argmaxinline_{j}\vect{f}(\vect{z})_j$.

\begin{definition}[certified robustness ($\ell_\infty$)]\label{def:certified_robustness}
    An input $\vect{z}$ is considered certifiably robust for a neural network $\vect{f}$ if the prediction remains unchanged for all perturbed versions: 
    \begin{equation*}
        \argmaxinline_j \vect f(\vect{z})_j = \argmaxinline_j \vect f(\Tilde{\vect{z}})_j,\quad \forall \Tilde{\vect{z}} \in \set{B}_\epsilon^\infty(\vect{z}).
    \end{equation*}
\end{definition}

Here, $\epsilon$ is the perturbation budget and $\Tilde{\vect{z}}$ is an element from the perturbation set based on the infinity norm:
$\set{B}_\epsilon^\infty(\vect{z}) = \{\Tilde{\vect{z}}\;\colon\norm{\vect{z} - \Tilde{\vect{z}}}_\infty \leq \epsilon\}$. 
If we cannot certify an input, it implies the existence of $\vect{z}^\prime \in \set{B}_\epsilon^\infty(\vect{z})$ for which $\argmaxinline_j\vect{f}(\vect{z})_j \neq \argmaxinline_j\vect{f(\vect{z}^\prime)}_j$. Any of these $\vect{z}^\prime$ instances are called adversarial examples.

The non-convex nature of the problem arises from the piece-wise linear characteristics of ReLU activation units. 
There are two approaches to address this issue: (i) model the ReLU activation with a binary variable or (ii) enclose the possible activation values $\vect{x}^{[i]}$ within a convex region.
The first approach results in a \textit{complete} formulation of the exact polytope, but the binary variables make the problem NP-hard \cite{tjeng2018evaluating}.
The second approach yields a \textit{convex} solution \cite{wong2018provable}.

\subsection{HQ-CRAN-DW}\label{sec:hq-cran-dw}

Here, we describe our algorithm designed to evaluate neural network robustness using the Dantzig-Wolfe decomposition. 
We build upon the formulation presented in \cite{franco2022quantum} and discuss the differences and adaptations for our approach.
To obtain a valid certificate, it is necessary to evaluate whether the network's prediction can be altered to any other possible class, as shown in \autoref{def:certified_robustness}.
However, w.l.o.g., we consider testing the difference between the initial predicted class and just one other class.
Thus, the original MILP problem instance is given by\footnote{The main distinctions involve substituting $\vect{z}$ with $\vect{x}$, $\vect{g}$ with $\vect{c}$, and $\vect{d}$ with $\vect{e}$. 
Additionally, the distinctions compared to the canonical form of \cref{eq:canonical_form} are that $\set{Y} = \{0, 1\}^{n_y}$ and $\set{X} = \left\{ \vect{x}\in\mathbb{R}^{n_x}\;\colon \matr{C}\vect{x} \geq \vect{e}\right\}$.
}:
\begin{equation}\label{eq:setup}
\min_{\vect{x},\vect{y}} \left\{\vect{c}^{\intercal}\vect{x} 
   \;\colon \matr{A}\vect{x} + \matr{B}\vect{y} \geq \vect{b}, \; \matr{C}\vect{x} \geq \vect{e} \right\},
\end{equation}
where $\vect{x}\in\mathbb{R}^{n_x}$ and $\vect{y}\in\{0, 1\}^{n_y}$ are the vectors of real and binary variables, respectively.

The master problem of Dantzig-Wolfe decomposition for \cref{eq:setup} is defined as:
\begin{subequations}\label{eq:master_setup}
\begin{align}
    \min_{\scriptsize \begin{array}{c} \lambda_i, \forall i \in \set{I}\\ \mu_j \forall j \in \set{J} \end{array}} 
    &\sum_{i \in \set{I}} (\vect{c}^\intercal \vect{x}) \lambda_i, \\
    \text{s.t.} \quad & \sum_{i \in \set{I}} (A \vect{x}^{(i)}) \lambda_i + \sum_{j \in \set{J}} (B \vect{y}^{(j)}) \vect{\mu}_j \geq \vect{b}, \\
    &\sum_{i \in \set{I}} \lambda_i = 1, \quad \lambda_i \geq 0,\,\forall i \in \set{I}, \\
    &\sum_{j \in \set{J}} \mu_j = 1, \quad \mu_j \geq 0,\,\forall j \in \set{J}, 
\end{align}    
\end{subequations}
where we omitted the integrality constraint on $\vect{y}$, as previously discussed in section~\ref{sec:dual}, it can be violated at a price of $\vectsym{\alpha}$~\cite{vanderbeck2010reformulation}.
The restricted version of \cref{eq:master_setup} is derived from the sets $\set{I^\prime} \subseteq \set{I}$ and $\set{J^\prime} \subseteq \set{J}$.
This restricted version is easier to solve and can provide initial solutions for the original master problem.
Subsequently, the \textit{real pricing} problem is given as: 
\begin{equation}\label{eq:sub_real_verification}
r = \min_{\vect{x}} \left\{ (\vect{c}^\intercal - \vectsym{\alpha}^{(t)}A) \vect{x}\;\colon \matr{C}\vect{x} \geq \vect{e}\right\}, 
\end{equation}
where $\vect{x} \in \mathbb{R}^{n_x}$ and the \textit{binary pricing} problem is given by:
\begin{equation}\label{eq:sub_binary_verification}
 p = \min_{\vect{y}} \{ - \vectsym{\alpha}^{(t)} \matr{B} \vect{y} \},
\end{equation}
where $\vect{y} \in \{0, 1\}^{n_y}$.
The Dantzig-Wolfe decomposition method iteratively solves the restricted master problem and the pricing problems until convergence is reached. 
Thus, given the original problem instance in terms of extreme points, we can state our main result.
\begin{theorem}
    Given a neural network $\vect{f}$ and an input $\vect{z}$, the solution of \cref{eq:master_setup} is a valid lower bound to the robustness verification problem of \cref{eq:setup}.
\end{theorem}

\begin{proof}
This is a direct consequence of \autoref{def:feasible_region}. 
The two sets of extreme points, $\set{P}_\set{X}=\left\{ \vect{x}^{(i)},\; \forall i \in \set{I} \right\}$ and $\set{P}_\set{Y}=\left\{ \vect{y}^{(j)},\; \forall j \in \set{J} \right\}$, are derived from \cref{eq:sub_real_verification} and \cref{eq:sub_binary_verification}, respectively. 
The set of linear constraints $\set{X} = \left\{ \vect{x}\in \mathbb{R}^{n_x}\;\colon \matr{C}\vect{x} \geq \vect{e} \right\}$ is satisfied through the optimality of \cref{eq:sub_real_verification}, while the set $\set{Y} = \{0, 1\}^{n_y}$ represents a binary instance.
Therefore, as long as the two sets include all extreme points, according to \autoref{def:lagrangian_dual}, the solution of \cref{eq:master_setup} provides a valid dual bound, which is equal to the convex relaxation of \cref{eq:setup}.

\end{proof}

\begin{algorithm}
\caption{HQ-CRAN-DW}\label{alg:hqcran}
 \begin{algorithmic}[1]
 \renewcommand{\algorithmicrequire}{\textbf{Input:}}
 \renewcommand{\algorithmicensure}{\textbf{Output:}}
 \REQUIRE $\vect{z}, \vect{f}(\vect{z}), \epsilon, T, \theta$
 \ENSURE  robust, not robust, or unknown \\
    \textit{Propagate Interval bounds}
    \STATE lower bound $\leftarrow$ CROWN-IBP from \cite{zhang2020towards}
    \STATE \textbf{if} lower bound $> 0$ \textbf{than} \textbf{return} robust \COMMENT{certified} \\
    \textit{Compute problem matrices}
    \STATE $\matr{A}, \matr{B}, \matr{C}, \vect{c}, \vect{b}, \vect{e} \leftarrow $ Alg. 1 from \cite{franco2022quantum}\\
    \textit{Find initial extreme point} 
    \STATE $\vect{x}^{(0)}, \vect{y}^{(0)} \leftarrow $ solve \cref{eq:setup} without objective 
    \STATE $\set{I}^\prime, \set{J}^\prime \leftarrow \vect{x}^{(0)}, \vect{y}^{(0)}$ \COMMENT{initialize extreme points sets}
    \STATE $\vectsym{\alpha}^{(0)}, \xi^{(0)}, \eta^{(0)} \leftarrow $ get dual from the relaxed version of \cref{eq:setup} without objective (i.e. $\vect{y} \in [0, 1]^{n_y}$). \\
    \FOR {each adversarial class}
        \STATE \textit{Iterate between the master and sub problems}
        \FOR {$t$ \textbf{in} $\{0, \dots, T\}$}
            % \textit{Solve pricing problems}
            \STATE $r, \vect{x}^{(t)} \leftarrow$ solve \cref{eq:sub_real_verification} with $\vectsym{\alpha}^{(t)}$ \COMMENT{classical}
            \STATE \textbf{if} $r < \xi^{(t)}$ \textbf{then} $\set{I}^\prime \leftarrow \vect{x}^{(t)}$
            \STATE $p, \vect{y}^{(t)} \leftarrow$ solve \cref{eq:sub_binary_verification} with $\vectsym{\alpha}^{(t)}$ \COMMENT{quantum}
            \STATE \textbf{if} $p < \eta^{(t)}$ \textbf{then} $\set{J}^\prime \leftarrow \vect{y}^{(t)}$
            \\
            \STATE $\varphi, \vectsym{\lambda}, \vectsym{\mu} \leftarrow$ solve \cref{eq:master_setup} with $\set{I}^\prime, \set{J}^\prime$
            % \STATE $\bar{\vect{x}} \leftarrow \sum_{i \in I} \vect{x}^{(i)}\lambda_i$ 
            % \STATE $\bar{\vect{y}} \leftarrow \sum_{j \in J} \vect{y}^{(j)}\mu_j$ 
            \STATE $\vectsym{\alpha}^{(t)}, \xi^{(t)}, \eta^{(t)} \leftarrow $ get dual solution from \cref{eq:master_setup} \\
            % \IF [Branching] {$\bar{\vect{y}} \in \mathbb{Q}$}
            %     \STATE add $\sum_{j \in \set{J}} \vect{y}^{(j)} \mu_j \leq \ceil{\bar{\vect{y}}}$ to \cref{eq:master_setup}
            % \ENDIF  \\
            \textit{Compute dual master objective}
            \STATE $\phi \leftarrow \max_{k \in\{0, \dots, t\}} - \vectsym{\alpha}^{(k)}\vect{b} - \xi^{(k)} - \eta^{(k)}$
            \STATE \textbf{if} $\varphi \leq 0$ \textbf{then} \textbf{return} not robust \COMMENT{adversary}
            \STATE \textbf{if} $\lvert \varphi - \phi \rvert \leq \theta$ \textbf{then} \textbf{break} \COMMENT{stopping criteria}
          \ENDFOR
          \STATE \textbf{if} $\phi \leq 0$ \textbf{then} \textbf{return} unknown \COMMENT{abstain}
    \ENDFOR
 \RETURN robust \COMMENT{certified}
 \end{algorithmic} 
\end{algorithm}

As observed, the binary pricing problem belongs to the class of unconstrained binary problems, which can be conveniently mapped into QUBO by considering $\texttt{diag}(-\vectsym{\alpha}^{(t)}\matr{B})$.
This eliminates the necessity for incorporating penalty terms through additional variables, which simplifies the overall process.
Consequently, the Dantzig-Wolfe decomposition is better suited to address the robustness verification problem of neural networks using QC.
However, in the convexification approach of \cref{eq:master_setup} integrality is required on $\vect{y}$ variables just as in the original \cref{eq:setup}.
Since we are not forcing the original problem to generate binary variables, the resulting solution $\bar{\vect{y}} = \sum_{\j\in\set{J}} \vect{y}^{(j)}\mu_j$ is a vector of real values between 0 and 1.
Therefore, an additional branching procedure is required to tight the convex relaxation.

In \autoref{alg:hqcran}, we present the HQ-CRAN-DW algorithm, which takes as inputs a neural network $\vect{f}$, a sample $\vect{z}$, a predefined $\epsilon$, a maximum number of steps T, and a predetermined gap $\theta$. 
The algorithm can yield one of three possible outcomes: certified robust, not robust, or unknown.
By employing CROWN-IBP~\cite{zhang2020towards}, the algorithm facilitates rapid convex propagation and estimation of bounds for $\vect{x}$.
If the global lower bound is positive, a certified sample can be returned immediately.

The procedure starts by determining an initial extreme point for each set from the solution of \cref{eq:setup} without the objective. 
This point acts as a feasible solution for all potential adversarial tests within the certification problem. 
Simultaneously, the convex relaxation of \cref{eq:setup} without the objective provides an initial dual solution, initiating the iterative process between the master and pricing problems. 
As new columns are constantly added, the algorithm refines and improves the master solution $\varphi$, converging towards the optimal one.

The algorithm terminates under three conditions: (i) when the difference between the master and sub is less than the threshold value, (ii) if the master objective is negative, indicating an adversary, and (iii) if the dual objective is below zero and the master is above zero, representing an unknown condition.
Otherwise, the dual master solution $\phi$ is positive for each of the adversarial class tested, the algorithm return a certified robust sample.
\section{Experimental Results}\label{sec:experiments}

In this section, we examine the application of Dantzig-Wolfe decomposition in the context of certifying the robustness of neural networks.
We report the details of the quantum hardware at the beginning of each experimental section.
On the classical side, we ran the algorithm on a server having 4xCPUs Intel(R) Xeon(R) E7-8867 v4 running at 2.40GHz for a total of 72/144 cores/threads. 

\subsection{Networks and Datasets}\label{app:settings}

We perform experiments using a single multilayer perceptron (MLP) neural network configuration: MLP-$2\text{x}[20]$. 
In this notation, MLP-$m\text{x}[n]$ represents a network with $m$ hidden layers and $n$ units per hidden layer. 
After each fully connected layer, ReLU functions are applied. 
Our model is trained on the MNIST dataset~\cite{lecun} for 20 epochs using a batch size of 128 in two distinct ways: (i) employing a standard loss function, and (ii) using adversarial training through Projected Gradient Descent (PGD), as described in \citet{madry2018towards}. 
For regularly trained models, we maintain the MLP-$2\text{x}[20]$ designation, while we refer to adversarially trained models as PGD-$2\text{x}[20]$. 
The clean test set accuracy for these networks is 95.62\% for MLP-$2\text{x}[20]$ and 86.73\% for PGD-$2\text{x}[20]$.
Adversarial training involves generating adversarial examples from an infinity norm ball surrounding the input, with a radius of $\epsilon = 0.01$.
\begin{figure}
    \centering
    \begin{subfigure}[b]{.45\textwidth}
        \includegraphics[width=\textwidth]{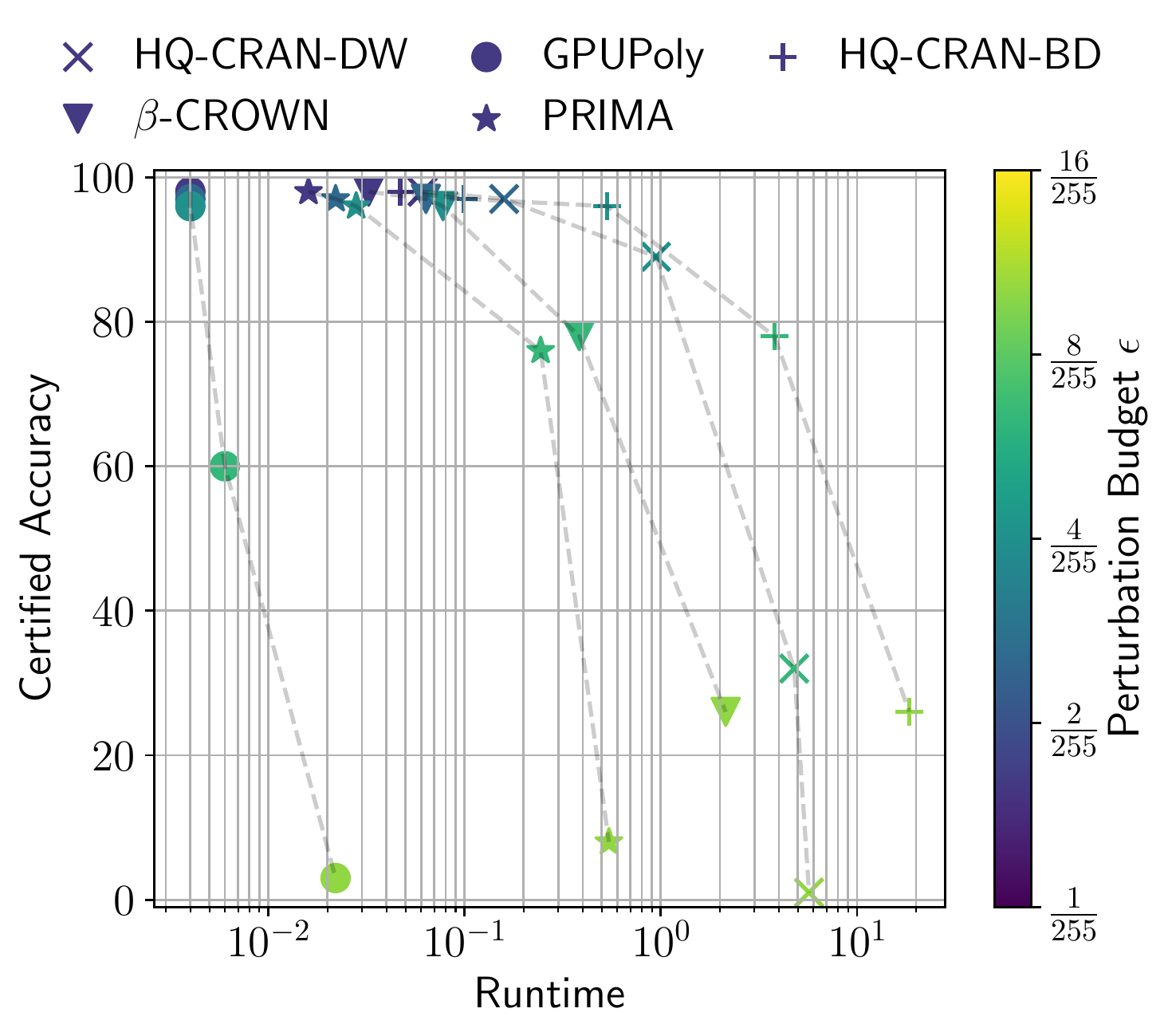}
        \caption{MLP-$2$x$[20]$}\label{fig:mpl_2_20}
    \end{subfigure}
    \begin{subfigure}[b]{0.45\textwidth}
        \includegraphics[width=\textwidth]{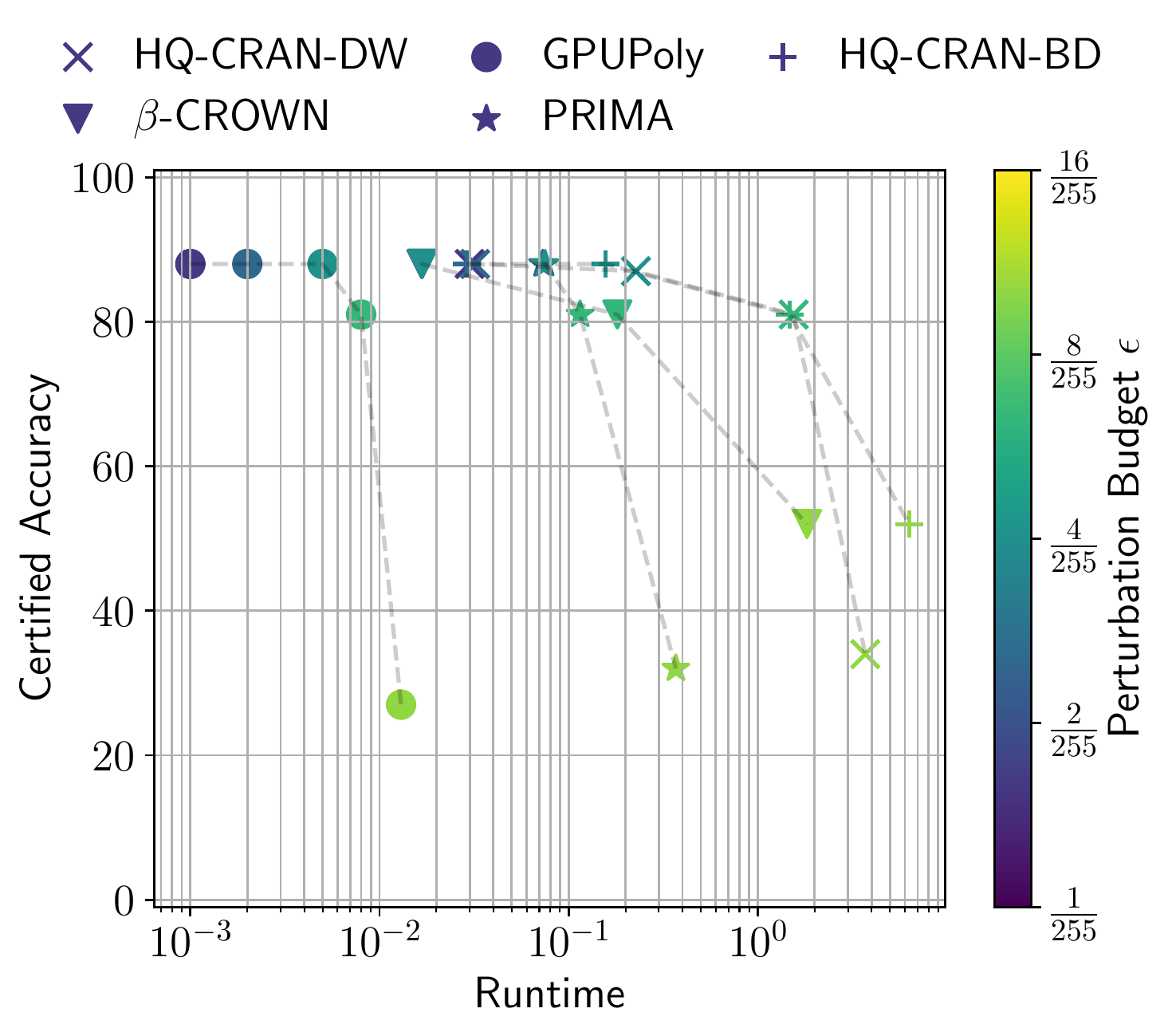}
        \caption{PGD-$2$x$[20]$}\label{fig:pgd_2_20}
    \end{subfigure}
    \caption{Average certified accuracy and runtime of various verification techniques including HQ-CRAN-DW/-BD for the initial 100 test MNIST samples on two MLP networks.}
    \label{fig:networks}
    \vspace{-1em}
\end{figure}

\subsection{Certified Accuracy}\label{sec:robustness}

In this section, we perform a comparison between our newly proposed variant, \textbf{HQ-CRAN-DW}, which employs Dantzig-Wolfe decomposition, and HQ-CRAN (v2)~\cite{franco2022quantum}, which utilizes Benders decomposition and is referred to as \textbf{HQ-CRAN-BD}.
Additionally, we compare the methods against the comprehensive verifier \textbf{$\beta$-CROWN}~\cite{wang2021beta}, as well as two convex verifiers \textbf{PRIMA}~\cite{muller2021prima} and \textbf{GPUPoly}~\cite{muller2021scaling}.
We assess HQ-CRAN-DW's empirical performance under optimal conditions, meaning that the master and sub-problems are resolved using the IBM ILOG CPLEX~\cite{cplex2009v12} software on a traditional computer. 
To ensure a fair comparison, we employ the IBP-CROWN~\cite{zhang2020towards} technique to propagate boundaries through the network, which is also utilized by $\beta$-CROWN. 
All methods are evaluated on the initial 100 samples from the MNIST test set, with adversarial budgets of $\epsilon \in \{\frac{1}{255}, \frac{2}{255}, \frac{4}{255}, \frac{8}{255}, \frac{16}{255}\}$.

We run HQ-CRAN-BD without the QUBO formulation for QC, meaning constraints are not relaxed with extra variables or incorporated into the objective via quadratic penalties.
Additionally, we consider the standard settings for HQCRAN-BD as described in~\cite{franco2022quantum}.

In \autoref{fig:networks}, we report the certified accuracy, as fraction of verified and correctly classified samples of all test samples, and runtime for two neural networks. 
In the context of HQ-CRAN-DW, the number of certified samples is similar than GPUPoly and PRIMA for $\epsilon$ values greater than $\nicefrac{8}{255}$ but lower than exact verifiers such as $\beta$-CROWN and HQ-CRAN-BD.
In terms of runtime, DW performs similarly than BD.
The limitations of HQ-CRAN-DW has to be related to the convex relaxation of the original MILP instance.
As noticing the number of certified samples reflects the same amount of convex verifiers.

\subsection{Simulated \& Quantum Annealing}\label{sec:annealing}

\begin{figure}
    \centering
    \begin{subfigure}[b]{.45\textwidth}
        \includegraphics[width=\textwidth]{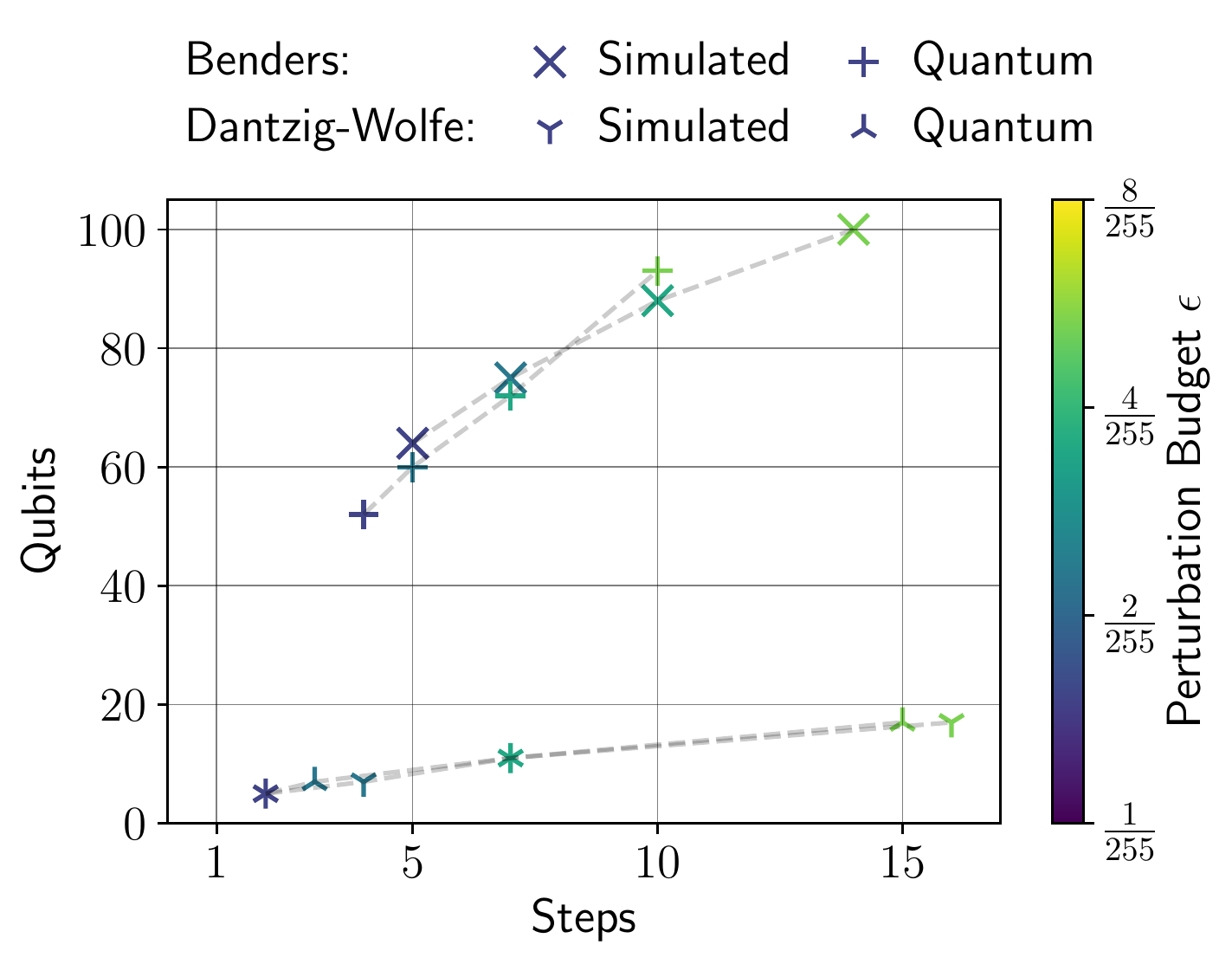}
        \vspace*{-7mm}
        \caption{MLP-$2$x$[20]$}\label{fig:mpl_2_20_annealing}
    \end{subfigure}
    \vfill
    \vspace{2mm}
    \begin{subfigure}[b]{0.45\textwidth}
        \includegraphics[width=\textwidth]{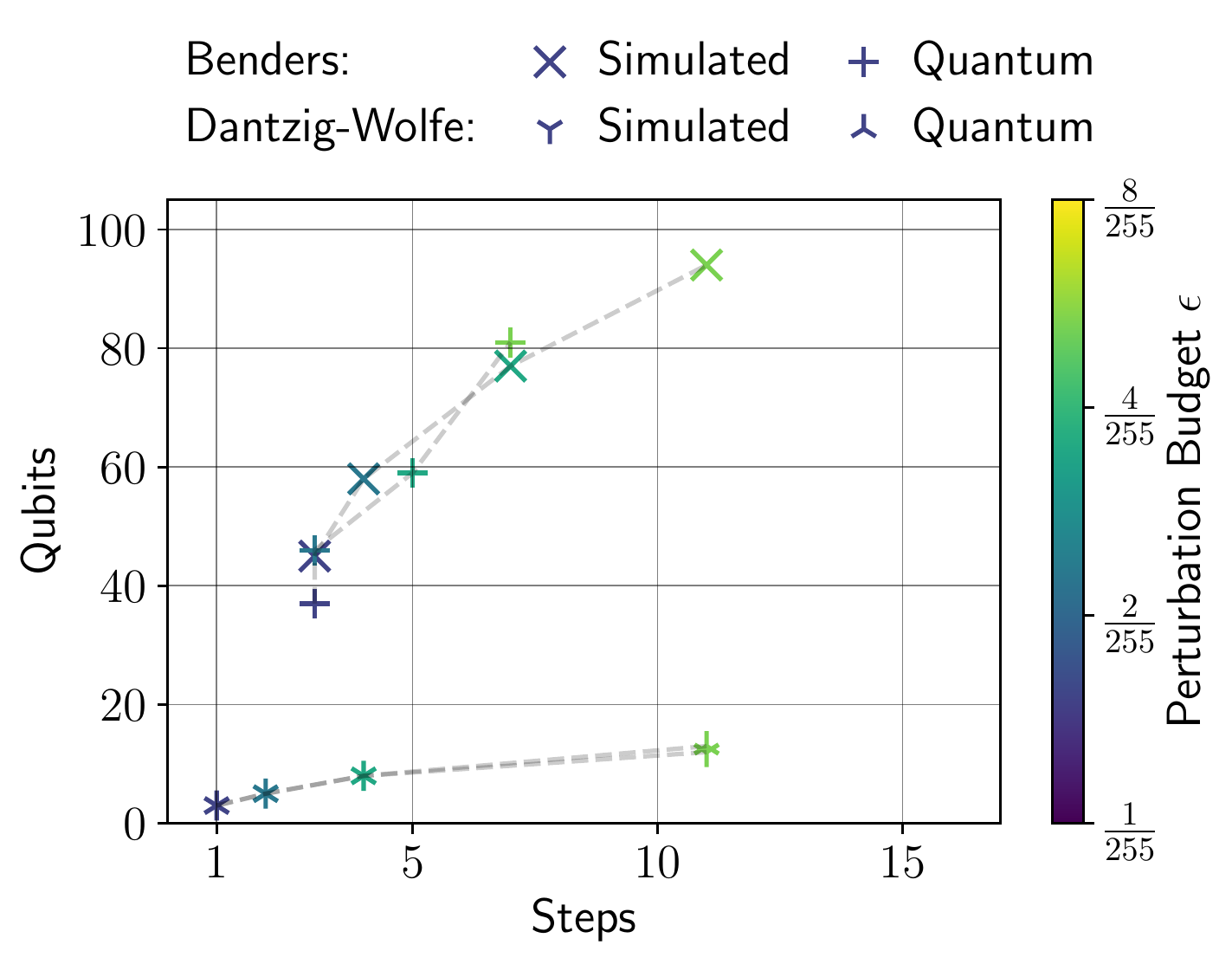}
        \vspace*{-7mm}
        \caption{PGD-$2$x$[20]$}\label{fig:pgd_2_20_annealing}
    \end{subfigure}
    \caption{Comparative analysis of Benders and Dantzig-Wolfe decompositions in the HQ-CRAN algorithm using simulated and quantum annealing on a 2-Layer MLP with 20 neurons per layer.}
    \label{fig:networks_annealing}
\vspace{-1em}
\end{figure}

Here, we proceed with our assessment by comparing simulated\footnote{Python dwave-neal v0.5.7, general Ising graph simulated annealing solver.} and quantum annealing.
The decision to select quantum annealing is linked to the size of the problem, particularly in relation to the evaluation of Benders decomposition.
On the quantum side, we access the D-Wave Advantage\textsuperscript{TM} system 4.1 constructed with 5760 qubits by D-Wave Leap\footnote{Quantum Cloud Service: \href{https://cloud.dwavesys.com/leap/}{https://cloud.dwavesys.com/leap/}}.

\subsubsection{Hyperparameters}
To embed the problem into a quantum annealer, the minor embedding problem must be solved. 
In our case, we employ the clique embedding strategy \cite{boothby2016fast}, which addresses the connectivity issue in a manner that any connection larger than the maximum available size (approximately 15 for Advantage) is managed with long chains between qubits. 
To reduce the chains, we prune connections below a specific threshold (5$\%$) before submitting the problem to the sampler, maintaining only interactions with values compatible with the sampler's precision. 
Both simulated and quantum annealing used 100 reads, while simulated annealing employed 50,000 sweeps.

In the context of Benders and Dantzig-Wolfe decompositions, the gap between the master and sub is set to $1$ and the maximum number of steps $T$ has been set to $15$ and $20$, respectively. 
The selection is connected to the fact that the number of qubits needed for Benders is a limiting factor when using the clique embedding, which sets the maximum size of logical qubits to $177$ on the Advantage~\cite{mcgeoch2021advantage}.
In the context of Benders, following the settings of~\cite{franco2022quantum}, the \textit{sub} problem boundaries $\overline{\vectsym{\alpha}}$ and $\overline{\vectsym{\beta}}$ are set to $5$. 
Additionally, the maximum size of the cuts set $\varphi$ has been set to $5$ and the penalty weights $w_a$ and $w_p$ were set to 0.1 and 0.01, respectively.

\subsubsection{Results}

In \autoref{fig:networks_annealing}, we show a comparison between Benders (BD) and Dantzig-Wolfe (DW) decomposition of the HQ-CRAN algorithm with simulated and quantum annealing.
We plot the average number of qubits and steps required to meet the predefined gap.
There is substantial difference between BD and DW in terms of qubits requirement.
In general the ratio\footnote{the ratio is calculated by dividing the average number of qubits needed for Dantzig-Wolfe by the average number of qubits needed for Benders and subtracting 1. To express the ratio as a percentage, the result is multiplied by 100.} of required qubits by DW is around 80\% (up to 90\%) less than BD.
Additionally, the number of steps needed by DW is lower for $\epsilon$ values smaller than $\nicefrac{8}{255}$.
The difference between quantum and simulated annealing is perceivable for larger problems, i.e. larger $\epsilon$ values.

\begin{table}[htbp]
\captionsetup{font=small}
\caption{Comparison of HQ-CRAN, with Benders (\textbf{BD})~\cite{franco2022quantum} and Dantzig-Wolfe (\textbf{DW}) decomposition with quantum annealing.
We run each algorithm on the first 100 samples of the MNIST test set.}
\begin{minipage}{\linewidth}
\begin{center}
\begin{footnotesize}
\begin{sc}
\begin{tabular}{llrrrrr}
\toprule
\multirow{2}{*}{Nets} & \multirow{2}{*}{$\epsilon$} &\multicolumn{3}{c}{Correct \& Certified $\uparrow$}
&\multicolumn{2}{c}{\# of Qubits $\downarrow$} \\
& &\multicolumn{1}{c}{CPLEX} &\multicolumn{1}{c}{BD} &\multicolumn{1}{c}{DW} &\multicolumn{1}{c}{BD} &\multicolumn{1}{c}{DW} \\
\midrule
\multirow{4}{*}{PGD}
    &$\nicefrac{1}{255}$ &$88\%$ &$61\%$ &$86\%$ &$37\pm17$ &$3\pm2$ \\
    &$\nicefrac{2}{255}$ &$88\%$ &$41\%$ &$77\%$ &$46\pm20$ &$5\pm2$ \\
    &$\nicefrac{4}{255}$ &$88\%$ &$20\%$ &$40\%$ &$59\pm21$ &$8\pm3$ \\
    &$\nicefrac{8}{255}$ &$81\%$ &$3\%$ &$3\%$   &$81\pm19$ &$13\pm3$ \\
\midrule
\multirow{4}{*}{MLP} 
    &$\nicefrac{1}{255}$ &$98\%$ &$49\%$ &$82\%$ &$52\pm20$ &$5\pm3$ \\
    &$\nicefrac{2}{255}$ &$97\%$ &$25\%$ &$70\%$ &$60\pm21$ &$7\pm3$ \\
    &$\nicefrac{4}{255}$ &$96\%$ &$7\%$ &$30\%$  &$72\pm24$ &$11\pm4$ \\
    &$\nicefrac{8}{255}$ &$78\%$ &$1\%$ &$2\%$   &$93\pm16$ &$17\pm4$ \\
\bottomrule
\end{tabular}
\end{sc}
\end{footnotesize}
\end{center}
\end{minipage}
\label{tab:quantum_annealing}
\end{table}

The numerical results of quantum annealing are presented in \autoref{tab:quantum_annealing}.
For a fair evaluation, alongside the number of qubits, we compare the percentage of correct and certified samples.
Correctness refers to the proportion of feasible solutions, indicating that the final master objective does not exceed the exact solution.
In the context of BD, the correct solution percentage declines as the adversarial perturbation budget $\epsilon$ rises. 
However, DW maintains consistent feasibility even with increasing values. 
As emphasized in \autoref{sec:comparison}, the master problem's solution in BD considerably influences the problem's feasibility, with simulated and quantum annealing potentially affecting solution quality and resulting in incorrect cuts and infeasible solutions. 
Conversely, DW solves the master problem classically, consistently satisfies coupling constraints and leading to feasible solutions. 
Nevertheless, the convexification of the original MILP formulation restricts the number of certified samples.

\subsection{Gate-Based vs. Annealing}

Here, we compare HQ-CRAN-DW running on quantum annealing or QAOA on a gate-base simulator.
In the first case, we consider the results of the quantum annelear from \autoref{sec:annealing}, while in the second case, we used the QAOA\footnote{Python library qiskit v0.36.0 \href{https://github.com/Qiskit/qiskit}{https://github.com/Qiskit/qiskit}} runtime program with a Aer\footnote{qiskit-aer v0.10.4 \href{https://github.com/Qiskit/qiskit-aer}{https://github.com/Qiskit/qiskit-aer}} simulator on classical hardware.
QAOA is considered with a depth of 5 and COBYLA~\cite{powell1994direct} as classical optimizer.

\begin{figure}
    \centering
    \begin{subfigure}[b]{.45\textwidth}
        \includegraphics[width=\textwidth]{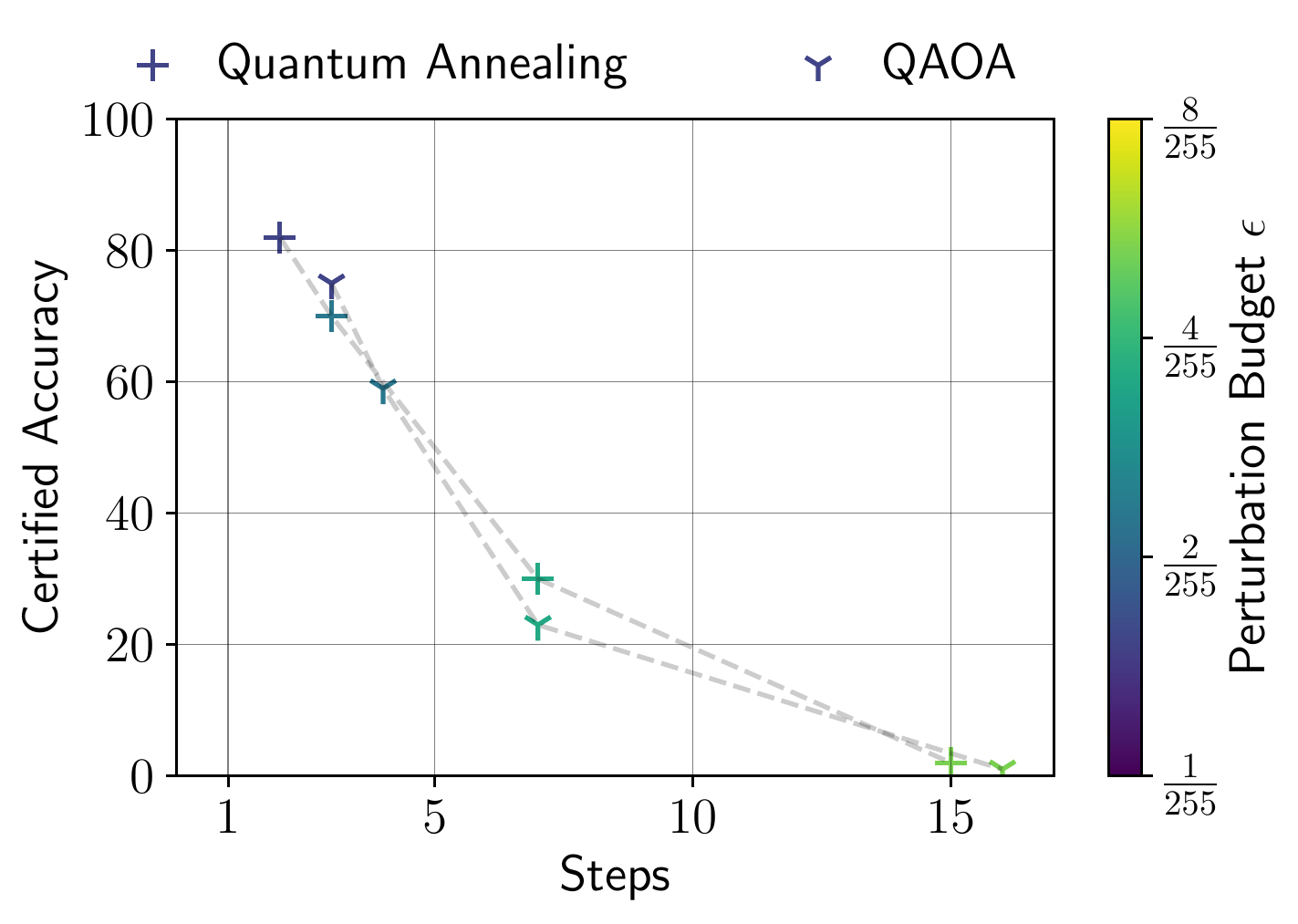}
        \vspace*{-7mm}
        \caption{MLP-$2$x$[20]$}\label{fig:mpl_2_20_simulated_qaoa}
    \end{subfigure}
    % \vfill
    % \vspace{2mm}
    % \begin{subfigure}[b]{0.45\textwidth}
    %     \includegraphics[width=\textwidth]{figures/qaoa/MLP2x20_qubits_qaoa.pdf}
    %     \vspace*{-7mm}
    %     \caption{PGD-$2$x$[20]$}\label{fig:pgd_2_20_simulated_qaoa}
    % \end{subfigure}
    \caption{Experimental analysis of HQ-CRAN-DW using quantum annealing and QAOA (Aer simulator) on the first 100 samples of the MNIST test set.}
    \label{fig:networks_qaoa}
    \vspace{-1em}
\end{figure}

In \autoref{fig:networks_qaoa}, we plot the average number of certified samples and steps with respect to increasing $\epsilon$ values.
It is worth noting that QAOA exhibits slightly inferior performance compared to quantum annealing, yielding fewer certified samples and requiring more steps for corresponding values.
Potentially, incorporating the CVaR metric~\cite{barkoutsos2020improving} could lead to substantial improvements in the results, enhancing the overall performance of QAOA.
However, considering that QAOA is essentially a trotterization of quantum annealing~\cite{farhi2014quantum}, these findings are in line with recent comparisons made on gate-based quantum hardware~\cite{pelofske2023quantum}.

\section{Discussion of results and limitations}\label{sec:discussion}

In this section, we discuss the outcomes of our experiments. 
A notable gap in runtime can be observed when comparing our proposed solution to classical neural network verifiers. 
This is due to the challenges faced in achieving high certified sample rates and the need for more interactions. 
These challenges stem from the basic implementation of the Dantzig-Wolfe decomposition, which introduces various computational issues~\cite{vanderbeck2005implementing}, such as the \textit{tailing-off effect} (slow convergence), the \textit{heading-in effect} (weak initial dual information), and the \textit{plateau effect}, which occurs when the master solution remains constant over multiple steps. 
However, several stabilization techniques have been developed to address these drawbacks. 
Specifically, integrating a more accurate branch-and-price procedure could lead to enhanced performance~\cite{vanderbeck2006generic}.

Despite the mentioned limitations, our method offers two main advancements in the field of hybrid verifiers for neural network robustness, both arising from Dantzig-Wolfe decomposition: (i) a fixed qubit number at each step, and (ii) the feasibility of the generated solutions. 
Additionally, the direct incorporation of a branch-and-price procedure would not compromise these benefits~\cite{ossorio2022optimization}.

\section{Conclusion}\label{sec:conclusion}

In this study, we have examined the complexity and qubit requirements of Benders and Dantzig-Wolfe decompositions for MILPs, with a particular focus on verifying the robustness of ReLU networks using QC. 
Since ReLU non-linearity can be expressed as a binary variable, the verification problem can be modeled as a MILP. 
Building on a previous approach~\cite{franco2022quantum}, we have proposed a Hybrid Quantum-Classical Robustness Analyzer for Neural Networks with Dantzig-Wolfe decomposition (HQ-CRAN-DW). 
Our finding show a reduction up to 90\% in qubits usage with respect to previous methods on quantum annealing and gate-based quantum computers.
Additionally, we demonstrate that the number of qubits required to solve Benders decomposition is exponentially large in the worst-case scenario, while it remains constant for Dantzig-Wolfe.

% \appendix
% \input{sections/appendix}

% \bibliography{refers/intro, refers/mip}
% \bibliographystyle{ieeetr}
\printbibliography

\end{document}